\documentclass[prb,preprint]{revtex4-1} 
\usepackage{amsmath}  
\usepackage{amsfonts} 
\usepackage{graphicx}
\begin{document}

\title{Easy method to establish the dispersion relation\\ of capillary waves on water jets}

\author{Wout M. Goesaert}
\email{w.m.goesaert@umail.leidenuniv.nl}
\author{Dr.ir. Paul S.W.M. Logman}
\email{logman@physics.leidenuniv.nl}

\affiliation{Leiden Institute of Physics, Niels Bohrweg 2, 2333 CA Leiden, The Netherlands}

\date{\today}

\begin{abstract}
A simple, intuitive, and low-cost setup for generating and measuring capillary waves is presented enabling a precise determination of the dispersion relation for a cylindrical water jet. By setting the phase velocity and measuring the wavelength of capillary waves directly, this method provides an intuitive way for students to understand the dispersion relation of a cylindrical water jet. The setup produced measurements of wavelength versus phase velocity over a broader range of values than earlier work. The resulting data are generally consistent with earlier results but show an error of up to 15\% at both the higher and the lower end of the measured range of wavelengths compared to the theoretical dispersion relation of cylindrical water jets. For the shorter wavelengths, the deviation is in the opposite direction from that of earlier work.
\end{abstract}

\maketitle

\section{Introduction}
One of the fundamental insights in understanding the complex behavior of fluids is the principle of dispersion: the difference in phase velocity of waves with different wavelengths. The dispersion relation, which relates wavelength to frequency in a medium, characterizes the amount and nature of dispersion and depends on both the material and the geometry through which the waves propagate. The understanding of dispersion is not only important in fluid dynamics; it also plays a role in solid state physics for the dispersion of phonons and was crucial in astronomy for a recent discovery of the missing baryonic matter problem. \cite{phonon}$^,$\cite{baryons}

\begin{figure}[h!]
\centering
\includegraphics[width=2.5in]{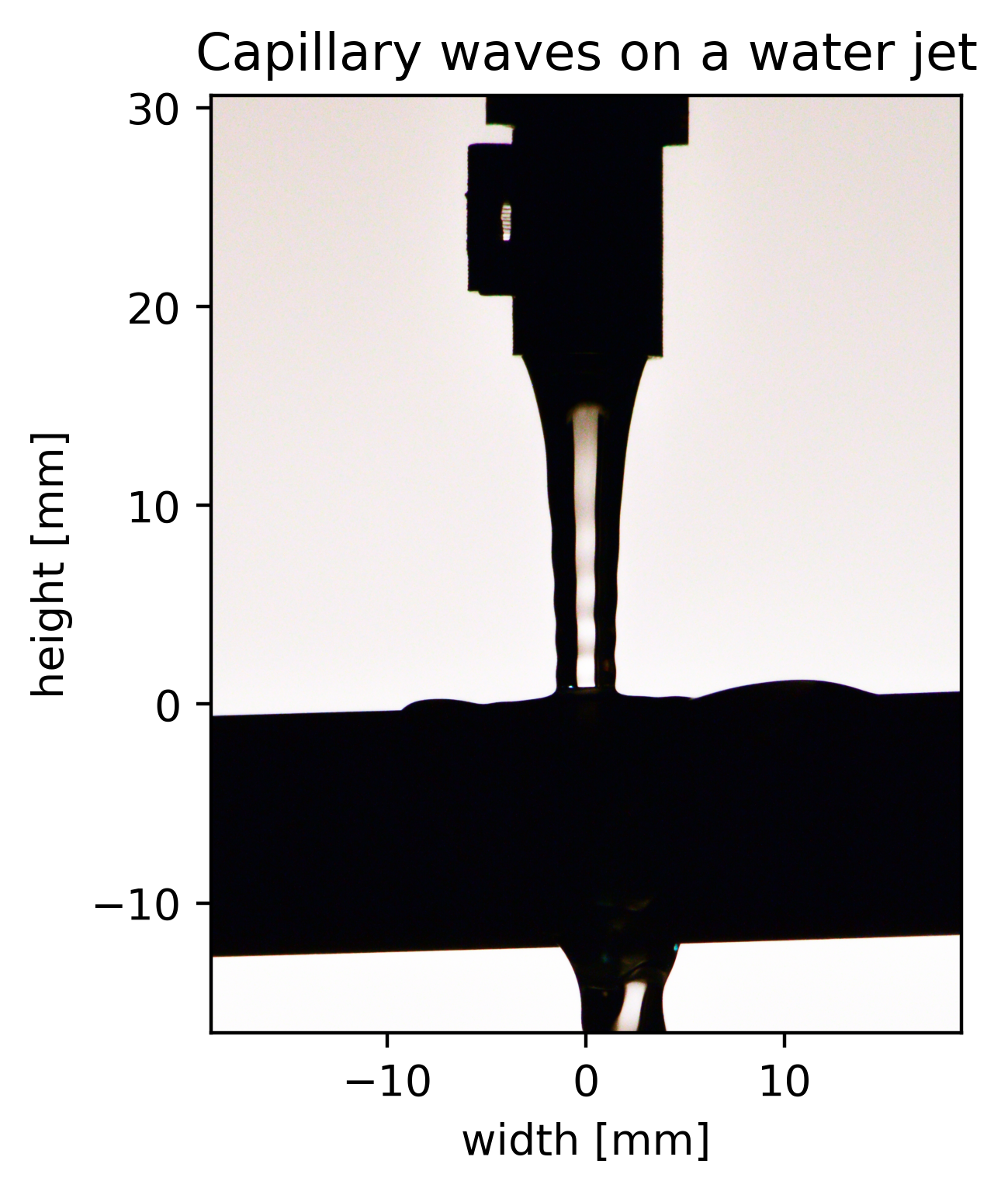}
\caption{Stationary waves appear on a laminar stream impinging on a rod. This wave pattern should not be confused with a standing wave, in which anti-nodes fluctuate over time (i.e. the jet would pulse over time).}
\label{fig:0}
\end{figure}

When a laminar cylindrical stream of water impinges upon an obstacle such as a solid surface or a body of water that obstructs the flow, a capillary wave pattern appears in the stream which is stationary relative to the laboratory (see Fig. \ref{fig:0}). The study of axisymmetrical capillary waves on water jets goes back to the work of Rayleigh.\cite{rayleigh} The effect gained public interest by being mentioned in the book "The Flying Circus of Physics"\cite{circus} and more recently caught the attention of a wider general public via YouTube\cite{stevemould}.

Capillary waves are surface waves caused by surface tension. In this case, they are the result of interference of upward traveling waves that originate at different times from the point of impingement. These waves disperse and interfere with the waves originating from earlier times. This interference is constructive only when the upward moving waves travel at a phase velocity that is equal to the downward stream velocity. The downward stream velocity depends on both gravity and surface tension and is difficult to calculate.\cite{goren,massalha} However, it can be easily determined experimentally by measuring the flow rate of water and the radius of the water jet.

Because of the interference of upward traveling waves, a completely stationary wave pattern in both space and time emerges (i.e. the jet and the wave in it appear frozen in time). As a result, otherwise small and fast traveling capillary waves become stationary relative to the laboratory and acquire a relatively large amplitude ($\sim 100\ \mathrm{\mu m}$) making them visible to the naked eye. The wavelength can then be measured directly from the stationary wave while the phase velocity can be set by varying the downward stream velocity. The frequency of the wave can be calculated from the values for phase velocity and wavelength, allowing for a measurement of both parameters in the dispersion relation.

Other researchers have done similar experiments in this field. For example, Zhu et al. created planar rather than cylindrical capillary waves using a capillary wave exciter.\cite{planarexperiment} They used laser diffraction to do their very precise wavelength measurements. Hancock \& Bush used pressure to create vertical water jets impinging on a deep fluid reservoir, thus creating cylindrical capillary waves.\cite{hancock} They made their wavelength measurements using digital photography. Both used professional equipment in their setup.

This experiment, however, uses only low cost and broadly available laboratory equipment. Water flow is created using gravity only. Cylindrical capillary waves are created by impinging the vertical water jet on a solid surface rather than a deep fluid reservoir. Like Hancock, wavelengths are measured using digital photography. This setup makes it possible to investigate cylindrical capillary waves at longer wavelengths than Zhu and Hancock did. It creates a cheap way for students to demonstrate and investigate the properties of the dispersion relation for a cylindrical water stream and to gain intuition about that relation. It opens up the possibility of further investigations on the influence of surface tension, viscosity, and the dispersion relation. This experiment can even be performed at home.

\section{Theoretical Background}\label{theory}
This section briefly summarizes the derivation of the dispersion relation for cylindrical water jets. For more details see the 1997 AJP paper by Sklavenites.\cite{Sklavenites}

The formation of stationary waves in water jets results from the fact that for capillary waves, which are driven by surface tension, short waves travel faster than long waves. For a flat surface of water, the planar model, this relation is given by:

\begin{equation}\label{eq:1}
    \omega^2 = \frac{\gamma}{\rho} k^3,
\end{equation}
where $\rho$ is the water density, $\gamma$ the surface tension, $k =\frac{2\pi}{\lambda}$ the wave number, and $\omega=v_{\mathrm{p}} k$ the angular frequency of the wave. Using the definitions of $\omega$ and $k$, this gives the relation between phase velocity $v_{\mathrm{p}}$ and wavelength $\lambda$:

\begin{equation}\label{eq:2}
    v_{\mathrm{p}}^2 = \frac{\gamma}{\rho} k = \frac{\gamma}{\rho} \frac{2 \pi}{\lambda}.
\end{equation}

However, this planar model only holds as long as $\lambda \ll R$ with $R$ the local stream radius. For longer wavelengths, the dispersion relation on a cylinder starts to deviate from that on a flat surface. The dispersion relation of the cylindrical model must then be used, which in the limit of vanishing viscous effects is given by:\cite{awati}

\begin{equation}\label{eq:3}
    \omega^2 = \frac{\gamma k}{\rho R^2} \frac{I_1(kR)}{I_0(kR)} \left(k^2 R^2 -1\right),
\end{equation}
which results in the following relation between phase velocity $v_{\mathrm{p}}$ and wavelength $\lambda$:

\begin{equation}\label{eq:3}
    v_{\mathrm{p}}^2 = \frac{\gamma \lambda}{2 \pi \rho R^2} \frac{I_1\left(\frac{2 \pi R}{\lambda}\right)}{I_0\left(\frac{2 \pi R}{\lambda}\right)} \left({\left(\frac{2 \pi R}{\lambda}\right)}^2 -1\right),
\end{equation}
where $I_1$ and $I_0$ are the modified Bessel functions of order 1 and 0 respectively. This analytical solution is based on the assumption of very small wave amplitude, allowing a linear approximation to be made.\cite{awati} The difference between the two models is largest for low $v_{\mathrm{p}}$ (i.e. long $\lambda$) and also for smaller surface tensions, as will be seen in Sect. \ref{result}. 

To see how the dispersion of waves results in the observed pattern, a simplified depiction of the effect is presented in appendix \ref{conceptualdepiction}. A description of the technical details that can be used to reproduce the visualizations in this depiction can be found in the corresponding online supplementary material.\cite{supplementary_material} This simulation follows the logic of Sklavenites with various approximations.\cite{Sklavenites}

The main point of this explanation is that because of constructive interference in capillary waves, effectively only waves with a characteristic wavelength $\lambda_{\mathrm{char}}$ will be observed. It also explains why $v_{\mathrm{phase}}=-v_{\mathrm{stream}}$.

Note that the approximation of no external force does not hold in the experiment presented here. Most importantly, the gravitational force will accelerate the water stream. The characteristic wavelength $\lambda_{\mathrm{char}}$, which is connected to the local stream velocity $v_{\mathrm{stream}}$ through the dispersion relation, therefore differs along the stream.

\section{Method}\label{setup}
Two variables are of interest for the measurement of the dispersion relation: $\lambda_{\mathrm{char}}$ and $v_{\mathrm{phase}}$. The characteristic wavelength $\lambda_{\mathrm{char}}$ of the capillary wave was measured directly. Since the waves are stationary, the phase velocity $v_{\mathrm{phase}}$ is equal in size to the local stream velocity $v_{\mathrm{stream}}$. To determine $v_{\mathrm{stream}}$ both the water flow rate $Q$ and the local radius $R$ of the stream were measured. The phase velocity then follows from:

\begin{equation}\label{eq:vel}
    v_{\mathrm{phase}}=\left|v_{\mathrm{stream}}\right|=\left|\frac{Q}{\pi R^2}\right|.
\end{equation}

\begin{figure}[h!]
\includegraphics[width=3.4in]{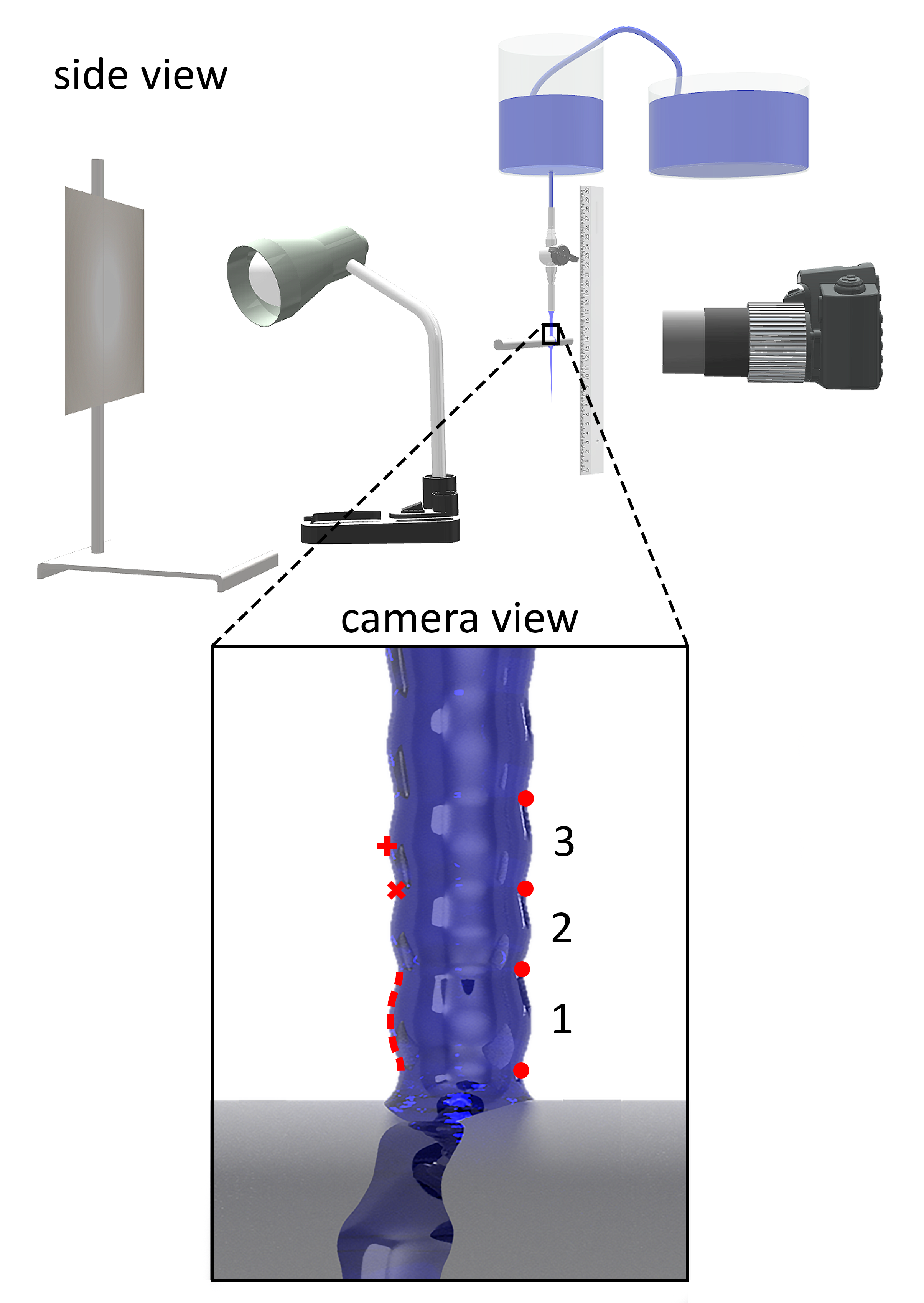}
\caption{(Color online) The experimental setup. Top: overview of the experimental setup with most of the structural elements omitted for clarity. The stream falls onto the rod between the camera and the illuminated screen so that the camera can capture the contour of the water flow. Bottom: silhouette of the stream as seen from the camera with a stream profile based on real data. On the left side of the stream, two methods of radius determination are illustrated using the radius of the stream averaged over a single wave (dashed line) and the mean of the radii at the wave extrema (cross and plus sign). On the right side, the three measured wavelengths are depicted.}
\label{fig:4}
\end{figure}

The capillary waves were created using a 1 L beaker hung from a tripod (see Fig. \ref{fig:4}). Connected to it via a siphon was a 2 L buffer with a large diameter such that the water level would only drop slowly. The beakers were filled with demineralized water. In our lab, beakers are available with a serrated glass hose barb connection at the bottom.\cite{glasswareshop} At the bottom of such a beaker, a rubber tube was attached with an on/off valve. Two Hoffman clamps were placed on the rubber tube: one to open and close and the other to regulate the flow rate. Because the end of the nozzle needed to be as level as possible, we created a nozzle using 3D printing and attached it to the end of the tube. The gcode-file to print the nozzle can be found in the corresponding online supplementary material.\cite{supplementary_material} The nozzle had an inner diameter of 5.78$\pm$0.01 mm though the exact diameter is not critical for the experiment. If 3D printing is not available, it should be possible to carefully mount a standard nozzle. A cylindrical rod was placed horizontally in the path of the stream below. The rod was connected via a vertical rod to a lab jack which could be lowered precisely to vary $v_{\mathrm{stream}}$ in small increments. A rod was utilized instead of a liquid bath as this fixes the point of impingement without the need of a drainage system. Below the setup, a bucket was placed to catch splashes.

The flow rate of the water stream was measured in separate trials from measuring the wavelengths and radii of the capillary waves. The first clamp on the rubber tube was set open and the second such that a stream of approximately 10 cm emerges before the water jet breaks up due to Plateau–Rayleigh instability. Hereafter, the second clamp was left untouched for the duration of the experiment.

The flow rate was measured using a measuring cylinder. The water level started at 1200 mL in the reservoir and was drained to 800 mL thereby dropping approximately 1.5 cm in height. Markers were put on the reservoir to indicate this range. The 20 times at which $20.0\pm0.5$ mL was drained were recorded using a timer that can record lap times. While keeping the water level within the indicated range, the water flow rate was seen to be constant throughout the rest of the experiment.

To double-check consistency in water flow rate, the above measurements were repeated several times within the marked range of water levels in the reservoir. This was done both before and after measuring the wavelengths and radii of the capillary waves.

For the precise measurement of the wavelengths and the radii of the capillary waves, a digital single-lens reflex camera ($f=200$ mm) was used. However, any reasonable camera would suffice. The camera was positioned 1 m from the nozzle. The experiment was performed in a dark room. In the background, 2 m from the camera lens, a strong light was directed at a white screen such that the silhouette of the stream against the white background could easily be seen.

For this experiment, python code was written to perform precise image analysis. Details on the python code can be found in the corresponding online supplementary material.\cite{supplementary_material}

The first three consecutive wavelengths as well as the mean radius over each wave were measured for each height of the rod (see Fig. \ref{fig:4}). This was done to investigate possible differences in the dispersion relation midstream compared to near the impingement point. At this point, the stream profile starts deviating notably from that of a cylinder.

In a class environment, students could instead measure the wavelength of the capillary waves from a digital photograph (e.g. using ImageJ).\cite{ImageJ} Measuring the mean stream radius precisely using such a program as well, may be asking too much of students. Instead, students can take the average of the peaks and troughs of a certain wave under the approximation that the wave profile is sinusoidal. An investigation of the accuracy of results using this alternate method proposed for students can be found in the corresponding online supplementary material.\cite{supplementary_material}

\section{Results}\label{result}
\begin{figure}[h!]
\centering
\includegraphics[width=3.4in]{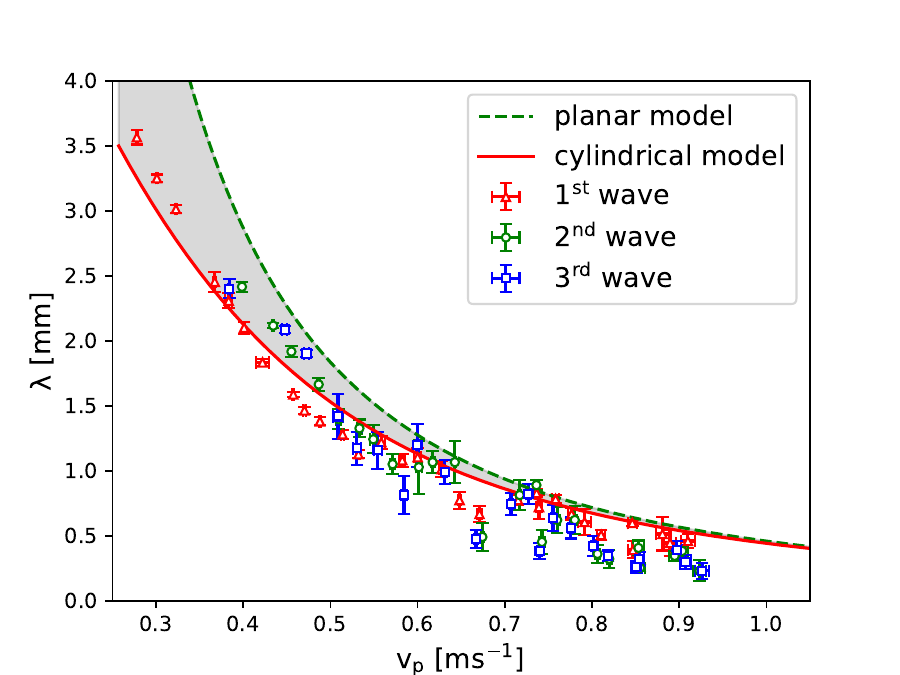}
\caption{(Color online) Comparison between the measured data and the models.  The cylindrical model (Eq. 4) predicts a relationship between the wavelength and phase velocity using the radius $R$ of the cylinder as a parameter.  The relationship for the smallest observed radius ($R = 1.08\ \mathrm{mm}$) is plotted as a solid red line, and the relationship for the largest possible radius ($R \rightarrow \infty$, which is equivalent to the planar model in Eq. 1) is plotted as a dashed green line. Therefore, the predictions for all possible radii must lie within the gray shaded area. We used the standard theoretical surface tension of $\gamma=$ 72 mN/m in both models. Smaller surface tensions for the model would result in smaller $v_{\mathrm{p}}$ at the same wavelength $\lambda$.}
\label{fig:5}
\end{figure}

The position of the rod was varied from $8.6$ mm to $46.1$ mm underneath the nozzle, resulting in wavelengths ranging from $3.57 \pm 0.06$ mm to $0.39 \pm 0.07$ mm as shown in Fig. \ref{fig:5}. For wavelengths longer than 3.57 mm the wave was less stable. For wavelengths smaller than 0.39 mm, the capillary waves became too small in amplitude to be measured accurately ($<0.02$ mm). For each height, 9 photo measurements were taken to determine both the radius R and the wavelength $\lambda$ of the wave.

The water flow in the nozzle ($Re \approx 750$) lies well within the laminar range which has a maximum Reynolds number of 1760.\cite{laminarflow1760} The water flow rate was constant throughout the experiment at $Q=3.43\pm0.04\ \mathrm{mL/s}$.

\begin{figure}[h!]
\centering
\includegraphics[width=3.4in]{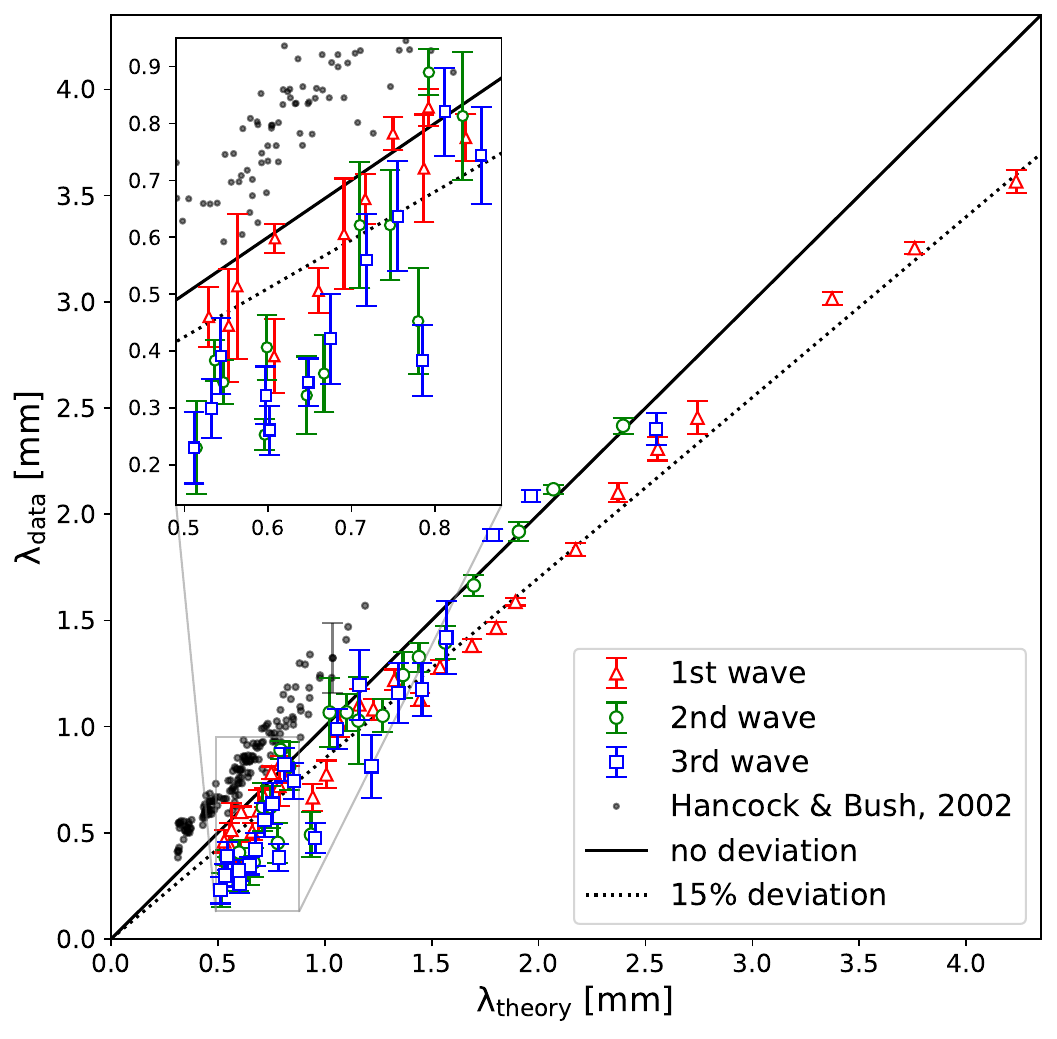}
\caption{(Color online) Discrepancy between data and the cylindrical model is plotted for the data in this study and that observed by Hancock \& Bush, where the flow rate was varied from $Q=1.861\ \mathrm{mL/s}$ to $Q=4.662\ \mathrm{mL/s}$. For the theoretical prediction, the local stream radii were used. The line at 15\% deviation is drawn as an indication of the magnitude of discrepancy at long wavelengths. Note that we make the comparison with the cylindrical model since its geometry most closely matches that of the stream. Again, we used a surface tension of $\gamma=$ 72 mN/m in this model. Smaller surface tensions in the model would result in a shorter wavelength $\lambda_{\mathrm{theory}}$ at the same wavelength $\lambda_{\mathrm{data}}$.}
\label{fig:6}
\end{figure}

As shown in Fig. \ref{fig:5}, the measured wavelength was observed to be consistently smaller than expected for velocities larger than $0.5\  \mathrm{m/s}$. The cylindrical model predicts the results most closely. The relation between the observed wavelength and the theoretically expected wavelength according to the cylindrical model is plotted in Fig. \ref{fig:6}. For comparison, the data from Hancock \& Bush is plotted in this figure as well.\cite{hancock}

\newpage

\section{Discussion}\label{discuss}
Using the presented setup, it was demonstrated that it is possible to measure the dispersion relation of capillary waves on a vertical water jet over a large range of wavelengths, including longer wavelengths than in previous research. A significant deviation with respect to both the planar and cylindrical theoretical model was found, especially at those longer wavelengths. By widening the nozzle and enlarging the reservoirs, the setup can be adapted to investigate the dispersion relation for even longer wavelengths.

Almost all measured wavelengths lie at lower phase velocities than their theoretical predicted phase velocities (see Fig. \ref{fig:5}). Water that has been exposed to air may have a lower surface tension than the theoretically used value of 72 mN/m. A lower surface tension would move the theoretical prediction more towards the measured data (smaller $v_{\mathrm{p}}$ at the same wavelength $\lambda$). However, the larger deviations from theory at both the longer and the shorter wavelengths indicate the presence of additional systematic deviations.

The (large) wavelength of the first wave was found to have a consistent deviation of around 15\% compared to the cylindrical model (see Fig. \ref{fig:6}). This could be due to the stream already starting to widen up near the impingement point, causing a deviation from a perfect cylinder, especially for wavelengths larger than 1.5 mm, for which the second and third waves don't show a discrepancy with the cylindrical model.

For wavelengths smaller than 1.5 mm, the wavelengths of all three waves were found to be smaller than both models predict. For the second and third waves, this discrepancy reaches around 50 \% at the shortest wavelengths. The exact reason for this discrepancy remains unclear. A possible explanation could be the finite amplitude effect.\cite{awati} Equation \ref{eq:3} is the solution of the hydrodynamic equations on a cylindrical column in the limit of waves with infinitesimal amplitude. This assumption doesn't hold in the phenomenon described in this paper, giving rise to a discrepancy called the finite amplitude effect. According to Denner, an amplitude larger than 0.05 $\lambda$ can result in deviations from theory that are larger than 2.5\%.\cite{denner} In our experiment, the amplitude ranged from 0.05 $\lambda$ for short wavelengths to 0.12 $\lambda$ for long wavelengths. Previous theoretical work on this finite amplitude effect suggests a decrease in theoretical wavelength.\cite{awati2}$^,$\cite{wang} However, previous experimental data from Hancock \& Bush and Sklavenites show a discrepancy towards longer wavelengths instead (see Fig. \ref{fig:6}).\cite{hancock}$^,$\cite{Sklavenites} This could be due to the difference in methodology. In their setup, the water jet fell onto a body of water instead of onto a rod and pressure was used to reach the preferred velocity. Further theoretical and experimental work is needed to investigate the finite amplitude effect and the difference between impingement on a rod and impingement on a body of water.

For the educational application of this experiment, however, the currently unexplained discrepancy is not a problem. When comparing the data with theory, students may even use the planar relation from Eq. \ref{eq:2}, as this approximation makes the analysis much simpler to perform. By finding a discrepancy in the data with respect to the model, students are offered a learning opportunity for dealing with results that do not (exactly) match their expectations.\cite{failure} Subsequently, they can practice the process of drawing correct conclusions while avoiding bias.\cite{bias}

As an extra challenge, students can largely use the same setup to study the effects of other variables on the dispersion relation as predicted by Eq. \ref{eq:3} (i.e. $\gamma$, $\rho$, $R$). This can be done by testing different concentrations of solubles and surfactants. An interesting follow-up research could be the relation between $\gamma$ and the observed discrepancy. If the discrepancy is due to the finite amplitude effect, it should diminish with decreasing surface tension and thus smaller amplitude. After all, when the surface tension becomes too low, the capillary waves will disappear altogether and the amplitude itself would become zero.\cite{hancock}

\section{Conclusion}\label{conclusion}
A simple setup was created for measuring the stationary capillary wave pattern on a cylindrical water jet. This low cost setup proves to be an effective way of demonstrating and determining the dispersion relation of capillary waves because the waves are stationary, large in amplitude, one-dimensional and the phase velocity can be easily calculated from the flow rate and the radius of the jet.

Furthermore, it proved possible to investigate the dispersion relation for longer wavelengths than in earlier research. It was also shown that the precision was sufficient to find deviations from theoretical predictions opening up the possibility to investigate these further. For example, a difference in wavelength was observed between the first and the two consecutive waves that could indicate an effect of the stream widening up right before the impingement point. Also, at short wavelengths, a downward discrepancy of 50\% was observed which may partly be due to the finite amplitude effect.

Because the utilized method differs from earlier experiments and yields opposite deviations, it is concluded that further theoretical and experimental work is needed to explain both observations. From this disparity between theory and data, a valuable learning opportunity arises for students by dealing with results that conflict with their expectations. It is also good to realize that providing first-year bachelor students, such as the first author, with open assignments hands them an opportunity to actively contribute to physics.

\begin{acknowledgments}
The authors would like to thank M. Hancock for providing his data and sharing his critical thoughts, M. Bergman for constructing a 3D model of the setup, L. Giomi, and T. Schmidt for their advice, and both internal and external reviewers for their valuable feedback which led to a substantial improvement of the setup and analysis.

In remembrance of our lab partner, contributor, and dear friend D.J. Remmelts.
\end{acknowledgments}

\section*{Conflict of interest}
The authors have no conflicts of interest to disclose.

\newpage
\appendix
\section{Depiction of dispersion of waves for first year bachelor students}\label{conceptualdepiction}
\begin{figure}[h!]
\centering
\includegraphics[width=3.7in]{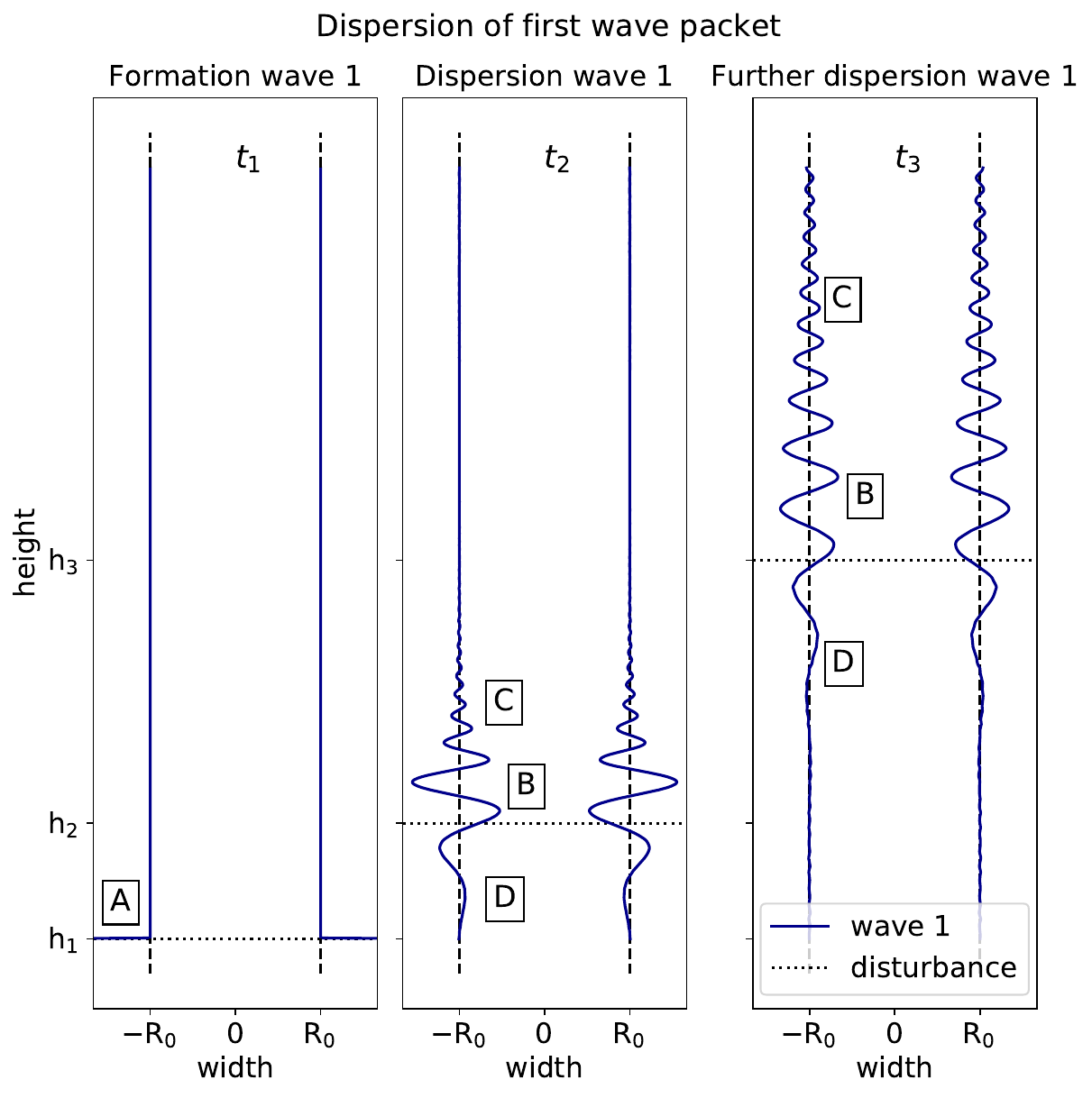}
\caption{The cross section of a cylindrical volume of stationary water is considered. On the left, the disturbance is depicted as a delta peak \boxed{\mathrm{A}} at $t_1$ and $h_1$. The delta peak is not part of the simulation but rather indicates a sudden disturbance with an unknown exact shape. The dispersion of the first wave packet is then shown at $t_2$ and $t_3$. Here, region \boxed{\mathrm{B}} indicates where $\lambda = \lambda_{\mathrm{char}}$, \boxed{\mathrm{C}} where $\lambda < \lambda_{\mathrm{char}}$ and \boxed{\mathrm{D}} where $\lambda > \lambda_{\mathrm{char}}$. The wave amplitude is exaggerated for illustrative purposes. Note that this is a 1D simulation that does not account for gravitational effects. Details on the simulation can be found in the corresponding online supplementary material.\cite{supplementary_material}}
\label{fig:1}
\end{figure}

Consider a vertical stationary cylinder of water with radius $R_0$ and no external forces (e.g. gravity) acting on it. Any internal friction is neglected because viscosity is negligible for water.\cite{Sklavenites} Suppose now that a disturbance were to move upward at a constant velocity $v_{\mathrm{stream}}$, constantly emitting waves in the process. We will first begin by simplifying this scenario, limiting ourselves to just two emitted waves and examining how these interact over time. Subsequently, we will progress to a more comprehensive depiction with a continuous emission of waves. The situation is evaluated at three random discrete time steps. At $t_1$, the disturbance point is located at height $h_1$ (see Fig. \ref{fig:1}). This disturbance, here depicted as a delta peak at point \boxed{\mathrm{A}}, produces capillary waves for a wide spectrum of wavelengths which travel outward as two wave packets in both directions from their point of origin. Only the upward moving wave packet is considered because the downward traveling waves will show to be irrelevant in our setup.

Because short wavelengths have a larger phase velocity $v_{\mathrm{p}}$ with respect to long wavelengths, the waves will have dispersed at $t_2$ such that long wavelengths lag behind. And at $t_3$, the wave packet will have dispersed even further. There exists some characteristic wavelength $\lambda_{\mathrm{char}}$ that corresponds to a phase velocity that is equal to the upward velocity of the disturbance: $v_{\mathrm{p,char}}=v_{\mathrm{stream}}$.  These waves travel with the disturbance, and are located at point \boxed{\mathrm{B}} in the figure. Shorter wavelengths ($\lambda < \lambda_{\mathrm{char}}$) are found higher up (at point \boxed{\mathrm{C}}) while longer waves ($\lambda > \lambda_{\mathrm{char}}$) lag behind (at point \boxed{\mathrm{D}}).

\begin{figure}[h!]
\centering
\includegraphics[width=3.7in]{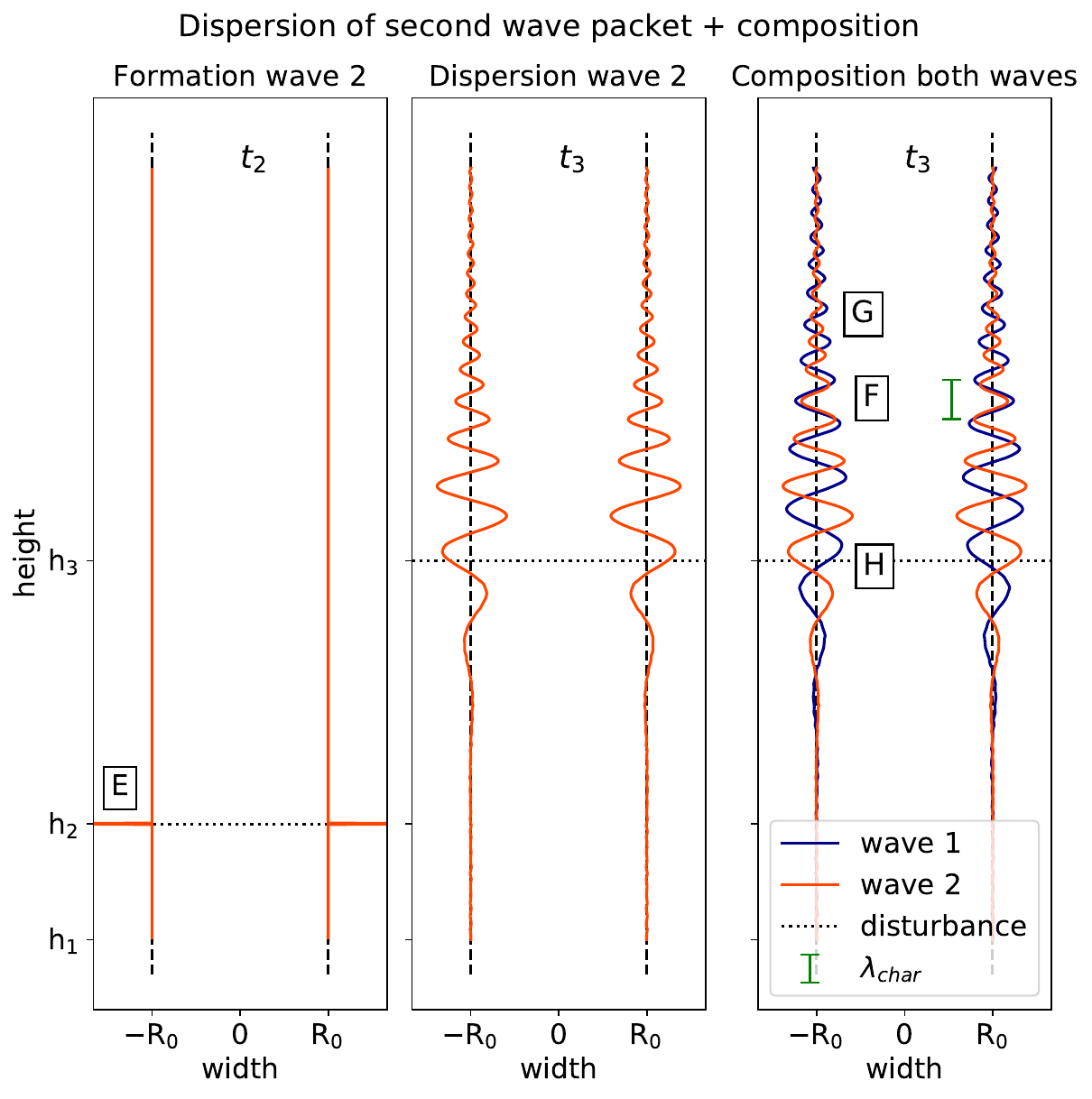}
\caption{The second wave packet is emitted at $t_2$ and $h_2$, again depicted on the left as a delta peak this time at point \boxed{\mathrm{E}}. The dispersion of the new wave packet is plotted at $t_3$ both on its own in the center and together with the first wave packet on the right. Region \boxed{\mathrm{F}} indicates where $\lambda = \lambda_{\mathrm{char}}$, \boxed{\mathrm{G}} where $\lambda < \lambda_{\mathrm{char}}$ and \boxed{\mathrm{H}} where $\lambda > \lambda_{\mathrm{char}}$. Again, the amplitude is exaggerated for illustrative purposes. Details on the simulation can be found in the corresponding online supplementary material.\cite{supplementary_material}}
\label{fig:2}
\end{figure}

Now consider a second identical set of waves which is emitted by the disturbance at $t_2$ when the disturbance is located at $h_2$ (point \boxed{\mathrm{E}}) (see Fig. \ref{fig:2}). On the third time step, $t_3$, the second wave packet has also dispersed in a similar manner. Again, waves with a characteristic wavelength $\lambda_{\mathrm{char}}$ move with the same phase velocity as the disturbance point. As a consequence, they have the same phase as the old set of waves and thus constructively interfere \boxed{\mathrm{F}}. However, the new set of waves has dispersed to a lesser degree than the first wave packet. As a result, the shorter wavelengths do not constructively interfere with each other \boxed{\mathrm{G}}. Similarly, wavelengths longer than $\lambda_{\mathrm{char}}$ also differ in phase and thus also do not constructively interfere \boxed{\mathrm{H}}.
\newpage

As can be seen in Fig. \ref{fig:2}, the point of constructive interference is located above the disturbance \boxed{\mathrm{F}}. This is because the group velocity $v_{\mathrm{g}}$, which can be thought of as the velocity of the region in which a particular wavelength is found, is larger than the phase velocity. For the planar model, given by Eq. \ref{eq:1}, the group velocity is equal to:

\begin{equation}\label{eq:3.5}
    v_{\mathrm{g}} = \frac{d \omega}{d k} = \frac{3}{2} \sqrt{\frac{\gamma}{\rho} k} = \frac{3}{2} v_{\mathrm{p}}.
\end{equation}

Likewise, it is also true that $v_{\mathrm{g}}=\frac{d \omega}{dk} > v_{\mathrm{p}}$ for the cylindrical model.\cite{awati}
Because of this fact, the narrow region of constructive interference (i.e. the region where $\lambda_{\mathrm{char}}$ is found) moves upward over time faster than the disturbance does.

\begin{figure}[h!]
\centering
\includegraphics[width=2.5in]{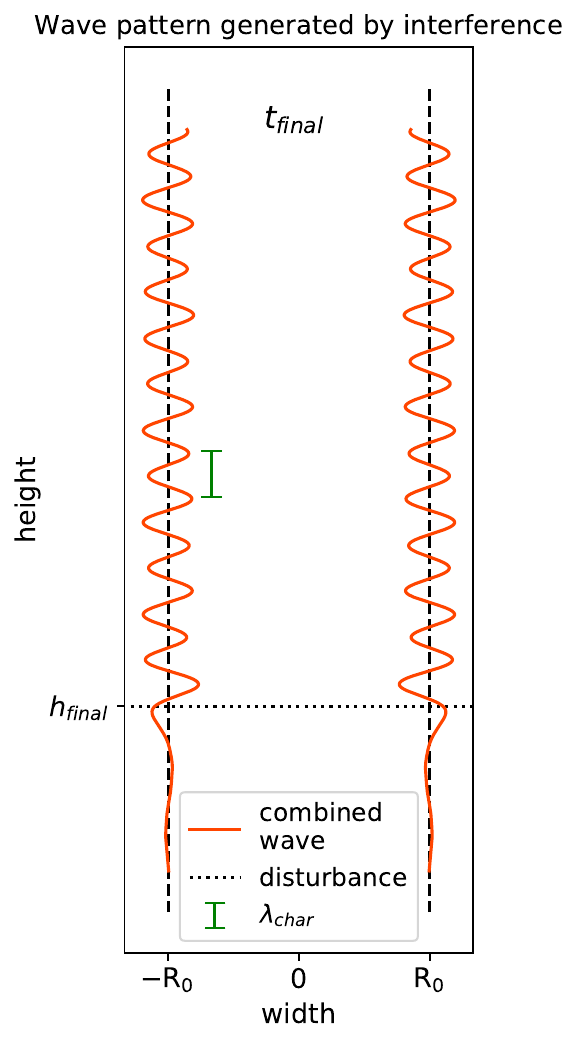}
\caption{The resulting wave pattern from the simulated 500 successive wave packets interfering at a much later time $t_{\mathrm{final}}$. Note that the amplitude is scaled and thus should not be compared to that in earlier figures. More details on the simulation can be found in the corresponding online supplementary material.\cite{supplementary_material}}
\label{fig:3}
\end{figure}

Finally, converting to the continuous production of waves therefore results in an interference wave pattern with $v_{\mathrm{g, char}} > v_{\mathrm{p, char}}=v_{\mathrm{stream}}$ moving upward. By considering a sum of successive wave packets generated at times before some $t_{\mathrm{final}}$ in the simulation, the observed wave pattern is recreated. A clean, stable and high amplitude wave with $\lambda = \lambda_{\mathrm{char}}$ which appears above the final impingement point at $h_{\mathrm{final}}$ is formed, as can be seen in Fig. \ref{fig:3}. The wave pattern appears stationary with respect to the disturbance because $v_{\mathrm{p, char}}=v_{\mathrm{stream}}$. Meanwhile, all other wavelengths are filtered out because they interfere destructively.

This capillary wave pattern is analogous to the Kelvin wake (the waves behind a ship) in which stationary gravity wave patterns appear behind an object traveling at constant velocity.\cite{Kelvin}

In the experiment, however, the water stream is falling onto a stationary disturbance. Therefore, we now convert to the laboratory's frame of reference which is simply that of the disturbance point such that waves moving with the same phase velocity as the obstruction appear stationary in the laboratory. Any downward traveling waves from the disturbance point are thus irrelevant, as no continuous stream is present underneath the obstruction. That is why only the upward traveling wave packets were considered here.

Summarizing, the observed stationary wave pattern is the result of constructive interference by capillary waves moving along with the obstruction.


\begin{thebibliography}{99}
\bibitem{supplementary_material}
The visualization simulation details, the Python code to recreate the simulation, the 3D-printing gcode-file of the nozzle (created with Ender 3), the Python code used for stream analysis and a description of two alternative (simpler) methods for students are shared at [url to be inserted by AIPP].
\bibitem{glasswareshop}
For those without access to a glassware shop, it may be possible to drill a hole in a clear plastic beaker and attach a hose barb connection using adhesive.
\end{thebibliography}
\end{document}


\title{Easy method to establish the dispersion relation\\ of capillary waves on water jets - Supplementary Material B.}

\author{Wout M. Goesaert Bsc.}
\email{w.m.goesaert@umail.leidenuniv.nl}
\author{Dr.ir. Paul S.W.M. Logman}
\email{logman@physics.leidenuniv.nl}

\affiliation{Leiden Institute of Physics, Niels Bohrweg 2, 2333 CA Leiden, The Netherlands}
\date{\today}

\maketitle

\section{Accuracy of two alternative (simpler) methods for students}\label{studentmethod}
To produce the results in this article, code was written to perform precise image analysis. This Python code can also be found in the online supplementary material.

However, undergraduate students may not have the skills to automatically process the images and distill the radius and wavelength of the capillary wave automatically from those images.

\begin{figure}[h!]
\centering
\includegraphics[width=4.5in]{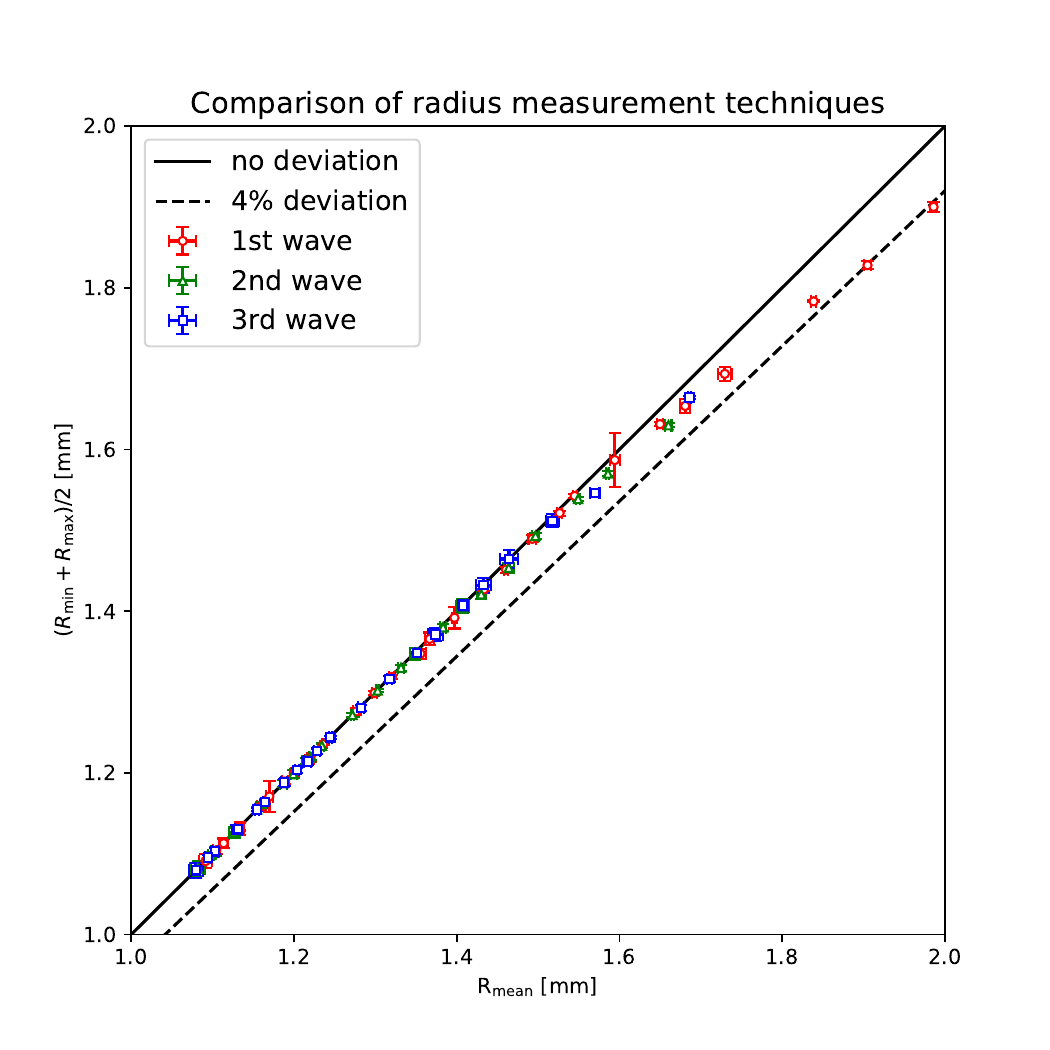}
\caption{Comparison of two methods for measuring the radius of the water stream over a wavelength. Each data point consists of 9 radius measurements. The small size of the introduced systematic error shows that students can use the easier to perform minimum-maximum method instead of taking the mean over the whole wave.}
\label{fig:meanvsminmax}
\end{figure}

Students may instead use an easier way to measure the radius of the jet wave to calculate the stream velocity $v_{stream}$ via the second equation in the paper. It does not matter whether students measure the radius of the jet wave via the mean of the minimum and the maximum of the wave or via the mean of the sinusoidal wave for most wavelengths (see the red markings on the second figure in the paper for a visualization of both methods). However, for long wavelengths, a deviation is introduced of maximally 4\% at $\lambda = 3.57\pm0.06$ mm (see Fig. \ref{fig:meanvsminmax}).

\begin{figure}[h!]
\centering
\includegraphics[width=4.5in]{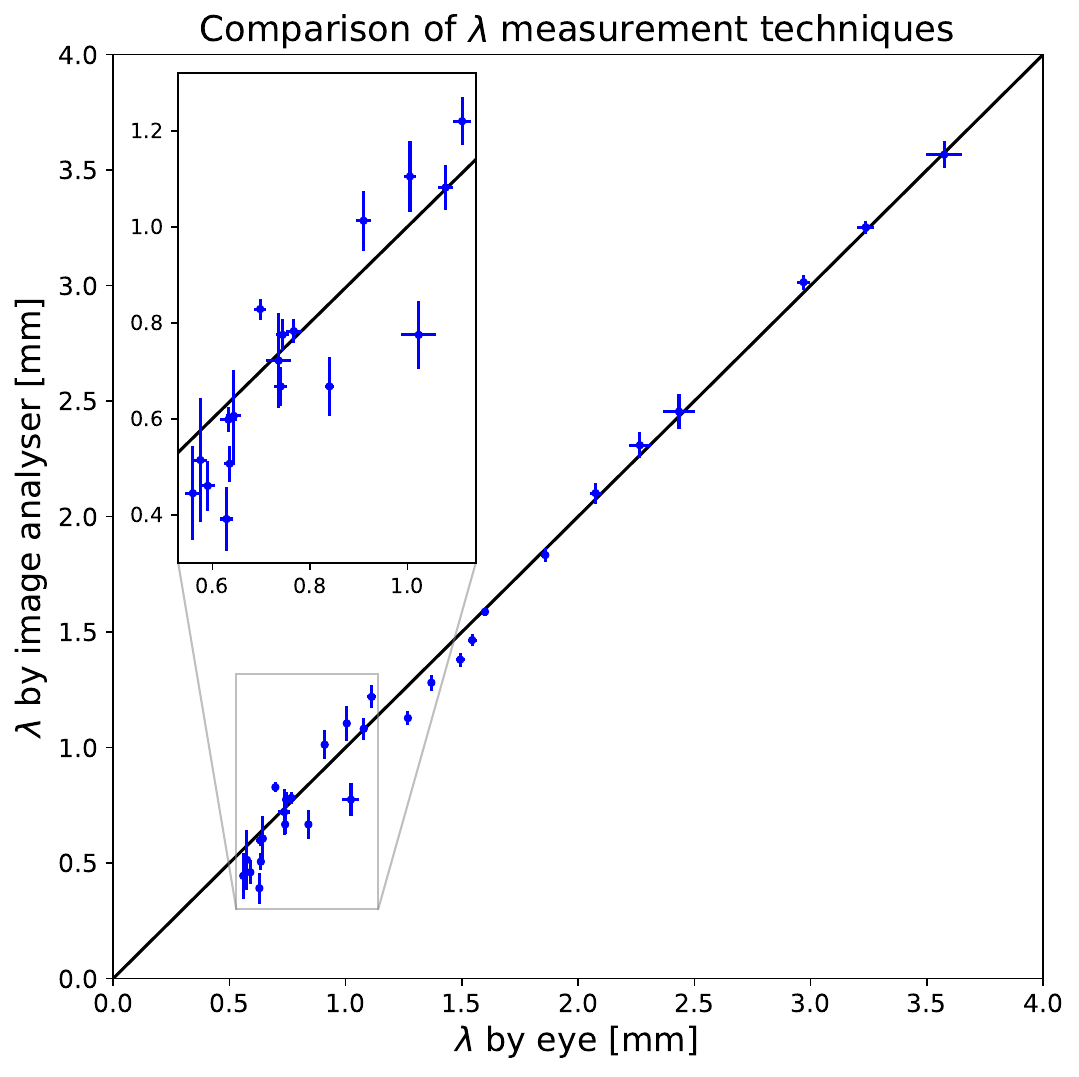}
\caption{Comparison of two methods for measuring the wavelength. Each data point consists of 18 measurements. A deviation is observed at the shortest wavelengths that is weakly significant.}
\label{fig:EyevsPython}
\end{figure}

Students may also choose to determine the wavelength of the capillary waves by means of directly analyzing digital photographs instead of writing a program to do this for them. This way, an error of up to 30 \% may be introduced for the shortest wavelengths (see Fig. \ref{fig:EyevsPython}). The errors for both the radius and the wavelength would however not account for the observed deviation from the theoretical dispersion relation in this article, since that deviation is larger. The easier measurements by directly analyzing digital photographs are found to result in smaller uncertainties. However, they are also more susceptible to biases.

